\documentclass[preprint]{aastex63}

\revised{January 29, 2021}
\accepted{January 30, 2021}
\shorttitle{STELLAR PARAMETRIZATION FOR M DWARFS}
\shortauthors{J.D. Li et al.}
\graphicspath{{./}{figures/}}

\usepackage{amsmath}
\usepackage{amssymb}
\begin{document}

\title{STELLAR PARAMETRIZATION OF LAMOST M DWARF STARS}

\correspondingauthor{Chao Liu}
\email{liuchao@nao.cas.cn}

\author[0000-0002-3651-5482]{Jiadong Li}
\affiliation{Key Laboratory of Space Astronomy and Technology, National Astronomical Observatories, CAS,  \\
Beijing 100101, China}
\affiliation{University of Chinese Academy of Sciences,  \\
Beijing 100049, China}

\author[0000-0002-1802-6917]{Chao Liu}
\affiliation{Key Laboratory of Space Astronomy and Technology, National Astronomical Observatories, CAS,  \\
Beijing 100101, China}
\affiliation{University of Chinese Academy of Sciences,  \\
Beijing 100049, China}

\author[0000-0002-6434-7201]{Bo Zhang}
\affiliation{Department of Astronomy, Beijing Normal University, \\
Beijing 100875, China}

\author[0000-0003-3347-7596]{Hao Tian}
\affiliation{Key Laboratory of Space Astronomy and Technology, National Astronomical Observatories, CAS,  \\
Beijing 100101, China}

\author{Dan Qiu}
\affiliation{China Three Gorges University, \\ Yichang 443002, China}
\affiliation{Center of Astronomy and Space Science Research, China Three Gorges University, \\ Yichang 443002, China}

\author{Haijun Tian}
\affiliation{China Three Gorges University, \\ Yichang 443002, China}
\affiliation{Center of Astronomy and Space Science Research, China Three Gorges University, \\ Yichang 443002, China}



\begin{abstract}
  M dwarf stars are the most common stars in the Galaxy, dominating the population of the Galaxy by numbers at faint magnitudes. Precise and accurate stellar parameters for M dwarfs are of crucial importance for many studies. However, the atmospheric parameters of M dwarf stars are difﬁcult to be determined. In this paper, we present a catalog of the spectroscopic stellar parameters ($T_{eff}$ and [M/H]) of $\sim$ 300,000 M dwarf stars observed by both LAMOST and {\it Gaia} using Stellar Label Machine (SLAM). We train a SLAM model using LAMOST spectra with APOGEE Data Release 16 (DR16) labels with $2800<T_{eff}< 4500$K and $-2<$[M/H]$<0.5$ dex. The SLAM $T_{eff}$ is in agreement to within $\sim 50$K compared to the previous study determined by APOGEE observation, and SLAM [M/H] agree within 0.12 dex compared to the APOGEE observation. We also set up a SLAM model trained by BT-Settl atmospheric model, with random uncertainties (in cross-validation) to 60K and agree within $\sim 90$K compared to previous study. 
\end{abstract}



\section{Introduction} \label{sec:intro}

In the Galaxy, M dwarfs are inherently faint objects, but dominate the faint magnitudes of the Galaxy by numbers, making up $\sim 70\%$ in our Galaxy \citep{2010AJ....139.2679B}. \cite{1999ApJ...519..802K} discovered spectral classes $L$ and $T$ using the Two Micron All Sky Survey (2MASS; \citealt{2006AJ....131.1163S}). $T$ type stars are completely comprised of brown dwarfs, while main-sequence (MS) stars earlier than M type are comprised of hydrogen-burning. M dwarf stars, which are at the MS end of the Hertzsprung-Russell (H-R) diagram, are in between the former and the later and show admixture features of both. 

The M dwarf stars have lifetimes much longer than the Hubble time \citep{2010AJ....139.2679B}, which makes them valuable for tracing the chemical and dynamical history of the Galaxy. Previous studies also used M dwarf stars to determine the initial mass function (IMF) \citep{2008AJ....136.1778C, 2010AJ....139.2679B} at low-mass end. Furthermore, M dwarfs are primary candidates for exoplanet searching \citep{2018A&A...609A.117T}. Accurate and precise parameters including effective temperatures and chemical compositions of the planet-host stars play a key role in looking for habitable exoplanets. 

The M dwarf stars are classified at wavelengths 6300 to 9000 \rm{\AA} \citep{boeshaar1976, kirk1992, boeshaar1985}. One of the main difficulties is that the prominent molecular absorption in the spectra of M dwarfs, are hard to predict by atmospheric models \citep{2015ApJ...804...64M}. Moreover, obtaining spectra with high quality for these faint objects is challenging. Generally, the measurement of equivalent widths (EWs) and synthesis are the classical and most common methods to derive stellar parameters. However, synthesis is the favorable method to measure the stellar parameters for cool stars since the EWs are difficult to be measured for their crowded absorption lines (reviewed by \citealt{2019ARA&A..57..571J}).

With the development of new facilities, large surveys such as the Sloan Digital Sky Survey, Apache Point Observatory Galactic Evolution Experiment (SDSS/APOGEE; \citealt{2017AJ....154...94M}), Transiting Exoplanet Survey Satellite \citep{2018AJ....155..180M} (\textit{TESS}) missions and the Large Sky Area Multi-Object Fiber Spectroscopic Telescope (LAMOST; \citealt{2012RAA....12.1197C, 2012RAA....12..735D, 2012RAA....12..723Z}) make incremental photometric and spectroscopic data of M dwarf stars. The LAMOST survey has provided nine million spectra in its Data Release 6 (DR6) at R $\sim 1800$, among which $\sim$ 600,000 spectra are M dwarf stars. However, these stars are lack of stellar parameters.

Many efforts have attempted to decode the effective temperatures and chemical abundances of M dwarfs from high-resolution spectra either in the optical or near-infrared (NIR) band \citep{2005MNRAS.356..963W, 2014A&A...564A..90R, 2018A&A...620A.180R, 2017ApJ...851...26V, 2019ApJ...871...63M}. The APOGEE Stellar Parameter and Chemical Abundances Pipeline (ASPCAP) measurements of $T_{eff}$ and metallicity \citep{2016AJ....151..144G} for M dwarfs have been determined with precisions of $\sim$ 100K and 0.18 dex, respectively \citep{2016MNRAS.460.2611S} by fitting with the atmospheric models. Furthermore, the APOGEE data of SDSS Data Release 16 (DR16) \citep{2020AJ....160..120J} use new atmospheric grids which can estimate effective temperatures down to 3000K. \citeauthor{2020ApJ...892...31B} (2020, hereafter B20) using \textit{The Cannon} \citep{2015ApJ...808...16N, 2017ApJ...836....5H} derived the effective temperature and metallicities for 5,875 APOGEE M dwarfs with 87 sources from \citeauthor{2015ApJ...804...64M} (2015, hereafter M15) as training dataset. M15 estimated effective temperatures by comparing spectra with the BT-Settl atmospheric models \citep{2013MSAIS..24..128A} and calibrated their results using stars with determinations from interferometry \citep{2012ApJ...757..112B, 2013ApJ...779..188M}. 

Furthermore, \cite{2018A&A...620A.180R} determined parameters of 45 M dwarfs using high-resolution H-band spectra by fitting BT-Settl model grids. \cite{2020arXiv201200915D} obtained stellar parameters of five M dwarf systems by fitting BT-Settl atmospheric and test current stellar models. \cite{2020AJ....159..193G} presented effective temperatures, radii, masses, and luminosities for 29,678 M dwarfs from LAMOST DR1 using \textit{The Cannon} with a typical uncertainty of $T_{eff}$ of $\sim$110 K. They used stellar labels from {\it TESS} Cool Dwarf Catalog \citep{2018AJ....155..180M}, in which $T_{eff}$ was determined from the color–$T_{eff}$ relations in M15. In other words, the effective temperatures of all previous studies rely on the BT-Settl atmospheric model. It is noted that the parameters derived by the previous works display substantial systematic errors, since different works used the different spectral regions and lines with various methods \citep{2019ARA&A..57..571J}. 

Recently, new techniques to derive stellar parameters with machine learning algorithms \citep{2019ApJ...879...69T, 2019ApJS..245...34X} have become efficient for large data of spectra survey. Data-driven methods were illustrated as promising solutions in cool star parameterization \citep{2019ARA&A..57..571J}. These novel methods have well performed in transferring the known information from training datasets to entire datasets.

 In this work, we build a data-driven model for LAMOST spectra based on Stellar LAbel Machine (SLAM; \citealt{2020ApJS..246....9Z}) trained by APOGEE stellar labels and BT-Settl model atmospheres and synthetic spectra \citep{2013MSAIS..24..128A} to estimate the entire dataset of LAMOST M dwarf stars.

This paper is organized as follows. In section \ref{sec:method}, we introduce how SLAM works. In section \ref{sec:data}, we describe M dwarf spectra selected from the LAMOST and \textit{Gaia} surveys as well as the training dataset from APOGEE survey and BT-Settl model. We then present the results in section \ref{sec:results} and make comparision with previous works. Section \ref{sec:discussion} raises discussions about the caveats of the results, and we assess the robustness and performance of the results and draw conclusions in section \ref{sec:discussion}.

\section{Method} \label{sec:method}
Stellar LAbel Machine (SLAM), developed by \cite{2020ApJS..246....9Z}, has shown good performance in determining the stellar labels of LAMOST DR5. It is a data-driven model based on Support Vector Regression (SVR) \citep{vapnik1997support}, which is a robust non-linear regression model. The data-driven method has been demonstrated as one of the most practical ways to measure the stellar parameters of M dwarfs. Additionally, LAMOST data is suitable for data-driven methods because the spectra of low resolution are hard to perform the standard methods of measuring EWs for parameters estimations of cool stars \citep{2019ARA&A..57..571J}. Meanwhile, the large quantity of LAMOST dataset demands to conduct fast data-driven methods.

\subsection{Support Vector Regression} \label{subsec:SVR}
Support-vector machine (SVM, also support-vector network; \citealt{cortes1995support}) is one of the most important supervised machine learning algorithms to be used for classification and regression. The regression algorithms of SVM, named support-vector regression (SVR), has been used in many astronomical studies, particularly in spectral data analysis \citep{2012MNRAS.426.2463L, 2014ApJ...790..110L, 2015ApJ...807....4L, 2015RAA....15.1137L}. 

\subsection{SLAM} \label{subsec:SLAM}
SLAM has three hyper-parameters, two of which are penalty level ($C$) and tube radius ($\epsilon$) coming from the genetic SVR algorithm. The third one ($\gamma$) indicates the width of the radial basis function (RBF), which is the kernel adopted by SLAM.

The architecture of SLAM consists of 3 steps:
\begin{itemize}
  \item [1.]\textbf{Pre-processing}. We normalize the spectra of the training data; in the mean time, we also standardize both stellar labels and spectral fluxes so that their mean is 0 and variance is 1; 
  \item [2.]\textbf{Training}. We train the SVR model with stellar parameters as independent variables and flux at given wavelength as dependent variable at each wavelength pixel using the training dataset; 
  \item [3.]\textbf{Prediction}. We apply the optimized SVR model to predict the stellar labels for observed spectra.  
\end{itemize}

To choose the best-fit hyper-parameters at each wavelength, which is defined as training procedure, SLAM minimizes the $k$-fold cross-validated mean squared error (CV-MSE), which is defined as
\begin{equation} \label{MSE}
  MSE_{j} = \frac{1}{m} \sum_{i=1}^{m}[f_j(\vec{\theta_{i}}) - f_{i,j}] ^ 2, 
\end{equation}
where $\vec{\theta_i} = (T_{eff,i}, \log{g}_i, [M/H]_i)$ denotes the stellar label vector of the $i$th star in the training data, $f_j(\vec{\theta_i})$ is the $j$th pixel of the training spectra as a function of $\vec{\theta_i}$. In prediction procedure, the posterior probability density function of $\vec{\theta}$ for an observed spectrum can be written as

\begin{equation}
p(\vec{\theta}| f_{obs}) \propto p(\vec{\theta}) \prod_{j=1}^n p(f_{j,obs}|\vec{\theta}),
\label{posterior}
\end{equation}

where $p(f_{j,obs}|\vec{\theta})$ is the likelihood of the spectral flux $f_{j,obs}$ varing with $\vec{\theta}$ based on the trained SVR model and $p(\vec{\theta})$ is the prior of $\vec{\theta}$. The best estimate of stellar labels can be found at the maximum of the posterior probability $p(\vec{\theta}| f_{obs})$. In practice, the logarithmic form of Eq. (\ref{posterior}) as in below is used by adopting a Gaussian likelihood:
\begin{equation}
  \begin{aligned}
    \ln p\left(\vec{\theta} \mid f_{obs}\right)=&-\frac{1}{2} \sum_{j=1}^{n} \frac{\left[f_{j, obs}-f_{j,model}(\vec{\theta})\right]^{2}}{\sigma_{j, obs}^{2}+\sigma_{j, model}(\vec{\theta})^{2}} \\
    &-\frac{1}{2} \sum_{j=1}^{n} \ln \left[2 \pi\left(\sigma_{j, obs}^{2}+\sigma_{j, model}(\vec{\theta})^{2}\right)\right], 
  \end{aligned}
\end{equation}
where $f_{j,obs}$ is the $j$th pixel of the observed spectrum, $f_{j,model}(\vec{\theta})$ is the SVR model-predicted spectral flux corresponding to the stellar label vector $\vec{\theta}$. $\sigma_{j,obs}$ is the uncertainty of the $j$th pixel of the observed spectrum, and $\sigma_{j,model}(\vec{\theta})$ is the uncertainty of the $j$th pixel of the model-predicted spectrum given the stellar labels $\vec{\theta}$. In practice, $\sigma_{j,model}(\vec{\theta})$ is replaced with CV-MSE$_j$, which is independent of $\vec{\theta}$. SLAM adopts Maximum Likelihood Estimation (MLE) with Levenberg-Marquardt (LM, \citealt{1978LNM...630..105M}) least square optimizer as the optimization method to derive the most likely $\vec{\theta}$ for an observed spectrum.

\section{Data} \label{sec:data}

\subsection{LAMOST Data} \label{subsec:LAMOST Data}
    LAMOST (Guo Shou Jing Telescope) is one of the most efficient spectroscopic survey telescopes providing 9,919,106 low-resolution (R$\sim$1800) optical spectra, among which 8,966,416 are stellar spectra, in its sixth data release (DR6) \citep{2012RAA....12.1197C, 2012RAA....12..735D, 2012RAA....12..723Z}. 607,142 spectra are published as the M dwarf catalog in LAMOST DR6.
    
    We firstly select stars from LAMOST DR6\footnote{\url{http://dr6.lamost.org}} M dwarf catalog \citep{2014AJ....147...33Y, 2019MNRAS.485.2167G} cross-matched with \textit{Gaia} DR2 \citep{2018A&A...616A..10G}. Then samples are selected using the following criteria to obtain both reliable \textit{Gaia} photometry ($G_{BP}-G_{RP}$ color and $G$-band magnitude), astrometry (parallax) and LAMOST spectra.

\begin{enumerate}
\item {\tt\string parallax / parallax error > 5};
\item {\tt\string phot\_bp\_mean\_flux/phot\_bp\_mean\_flux\_error} $>$ 20,\\
    {\tt\string phot\_rp\_mean$\_$flux/phot$\_$rp$\_$mean$\_$flux$\_$error} $>$ 20,\\
    and {\tt\string phot$\_$g$\_$mean$\_$flux/phot$\_$g$\_$mean$\_$flux$\_$error} $>$ 20;
\textbf{\item {\tt\string ruwe < 1.4}}
\item signal-to-noise ratio (SNR) at $i$-band of LAMOST spectra is larger than 5.
\end{enumerate}

\begin{figure}[hbt!]
\centering
\includegraphics[width=9cm]{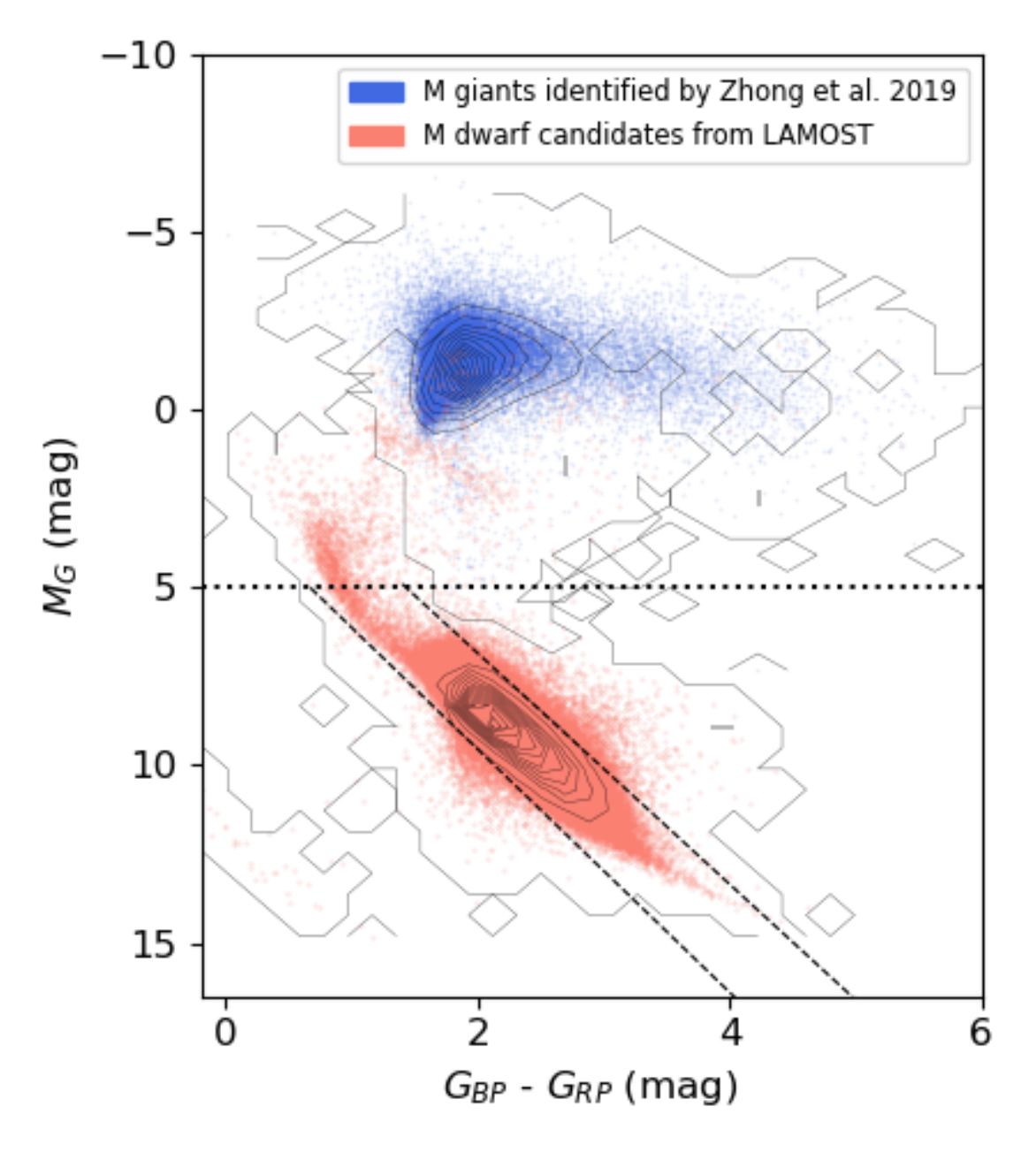}
\caption{The panel shows the color-magnitude diagram, i.e. $G$-band absolute magnitude ($M_G$) versus $G_{BP}-G_{RP}$, of selected M dwarf samples compared with M giants from \cite{2019ApJS..244....8Z}. The red stars represent the M dwarfs candidates from LAMOST survey, while the blue stars denote the M giants samples from \cite{2019ApJS..244....8Z}. The horizontal dotted line represents the $M_G$ equals to 5. The coordinates of the selected quadrangles enclosed by the dotted line and two dashed lines are [(0.7, 5), (1.4, 5), (4, 16.5), (5, 16.5)]. 
\label{fig:hrd_cut}}
\end{figure}

The criteria 1-3 aim to select stars with both accurate photometry and astrometry. {\tt\string ruwe} is the re-normalised unit-weight error which measures astrometric goodness-of-fit, crieria 3 is to select stars with small re-normalised unit-weight error. Criteria 4 aims to select spectra with clear stellar spectral signature. Similar to B20, criteria 5 aims to select the main-sequence M dwarfs as shown in Fig. \ref{fig:hrd_cut}. 

We display the selected M dwarf samples in H-R diagrams (Fig. \ref{fig:hrd_cut}). The $G$-band absolute magnitude is estimated from the Bayesian distance from \cite{Bailer-Jones2018}. To investigate the contamination of M giant stars, we draw 35,382 M giants on Fig. \ref{fig:hrd_cut} \citep{2019ApJS..244....8Z}. We select $M_G + A_G \le 5$ to remove the contaminations. Some of the M dwarf stars are not located at the main-sequence, but below and above the main-sequence. There are $\sim 7,000$ stars on the top side of the quadrangle which are likely pre-main sequence stars or binaries. And about 1,000 stars on the bottom side might be white dwarfs - MS binaries. We remove these stars and only select 379,258 M dwarf samples fell into the quadrangle for the estimation of the stellar parameters. Most of M dwarf stars are located within a few hundreds pc. Therefore, the interstellar extinction of M dwarfs are mostly very low. There may be a few stars with large extinction and thus location beyond the selection area. These stars may be removed mistakenly.

\subsection{Training Dataset} \label{sec:training dataset}
Since SLAM is a data-driven model, which assumes stellar labels as ground truth, reliable stellar parameters of training datasets are needed. To date, various methods and training datasets have been developed and introduced to measure labels (stellar parameters) for M dwarfs. In this work, we use APOGEE stellar parameters and BT-Settl synthetic spectra, respectively, as the training dataset separately described in the following two subsections \ref{subsec:ASPCAP} and \ref{subsec:BT-Settl}.

\subsubsection{APOGEE labels as training data} \label{subsec:ASPCAP}
\cite{2016AJ....151..144G} presented the ASPCAP, which fits observed near-infrared spectra to synthetic spectra made with the code FERRE \citep{2006ApJ...636..804A}. The measurement of $T_{eff}$ for M dwarfs reach the precision of 100K between $3550 < T_{eff} < 4200$K, the \textbf {mean} precision of ASPCAP metallicities is 0.18 dex between $-1.0<$ [M/H] $<0.2$ \citep{2016MNRAS.460.2611S}. For APOGEE DR16, \cite{2020AJ....160..120J} used the new MARCS stellar atmospheric models which is continuous from 3000 to 4000K for $T_{eff}$. The stellar parameters of DR16 enhance significantly for cool stars with $T_{eff} <$ 3500K, avoiding discontinuities in ASPCAP at 3500K. As for the metallicity of DR16, the comparison with six well-studied open clusters shows a faint difference of 0.004 dex \citep{2020AJ....159..199D, 2020AJ....160..120J}.

We cross-match our selected M dwarfs APOGEE DR16 catalog\footnote{\url{https://www.sdss.org/dr16/irspec/aspcap/}} and obtain about 4,317 common stars with LAMOST spectra and APOGEE labels as training data. We further select 3,785 samples using the following criteria: 
\begin{itemize}
    \item[1.] $2800 < T_{eff} < 4500$K;
    \item[2.] $T_{eff}$ uncertainty smaller than 100K;
    \item[3.] $-2<[M/H]<0.5$ dex;
    \item[4.]  [M/H] uncertainty smaller than 0.1 dex;
    \item[5.] $\log{g} >4$ dex. 
\end{itemize}

\begin{figure*}[hbt!]
\plotone{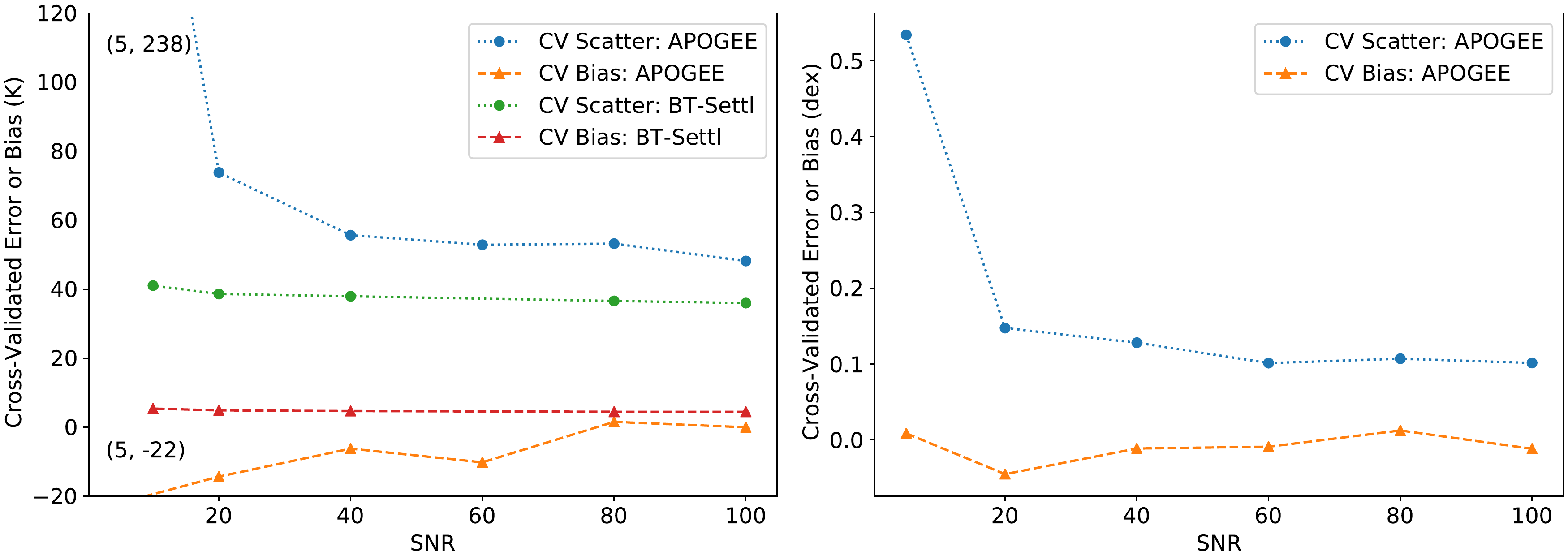}
\caption{The figure displays how the cross-validation (CV) errors of stellar labels change with the signal-to-noise ratio (SNR). In both panels, the dotted lines represent the CV-scatter, while the dashed lines denote the CV-bias. Note that the SNR for APOGEE labeled LAMOST spectra is the SNR in the $i$-band (SNR$_i$), the SNR of the BT-Settl synthetic spectra is added artificially. Clearly, all of the CV errors decrease as the SNR increases. When SNR$_i > 100$, the typical CV-scatters of $T_{eff}$ and [M/H] are 50K, 0.10 dex for the ASPCAP labels, respectively. The typical CV-scatter for the BT-Settl labels is 40K when SNR $>100$.
}
\label{fig:cv_error}
\end{figure*}

\subsubsection{BT-Settl as training data} \label{subsec:BT-Settl}
Another independent training dataset is BT-Settl spectra. Unlike the empirical spectra with APOGEE stellar labels, the BT-Settl model atmospheres and synthetic spectra \citep{2013MSAIS..24..128A} are computed by solving the radiative transfer using the Mixing Length Theory \citep{1958ZA.....46..108B}. The BT-Settl model can be used to determine the parameters from moderately active very-low-mass stars (VLMs), brown dwarfs to planetary-mass objects. The BT-Settl model would be a useful supplement and extend the effective temperature lower than 3000K. Since a data-driven method is more difficult to be trained at the edge of training labels, we finally use $T_{eff}$ range from 2200K to 7000K as the training labels.

\section{Results} \label{sec:results}

\begin{figure*}[hbt!]
  \centering
  \plotone{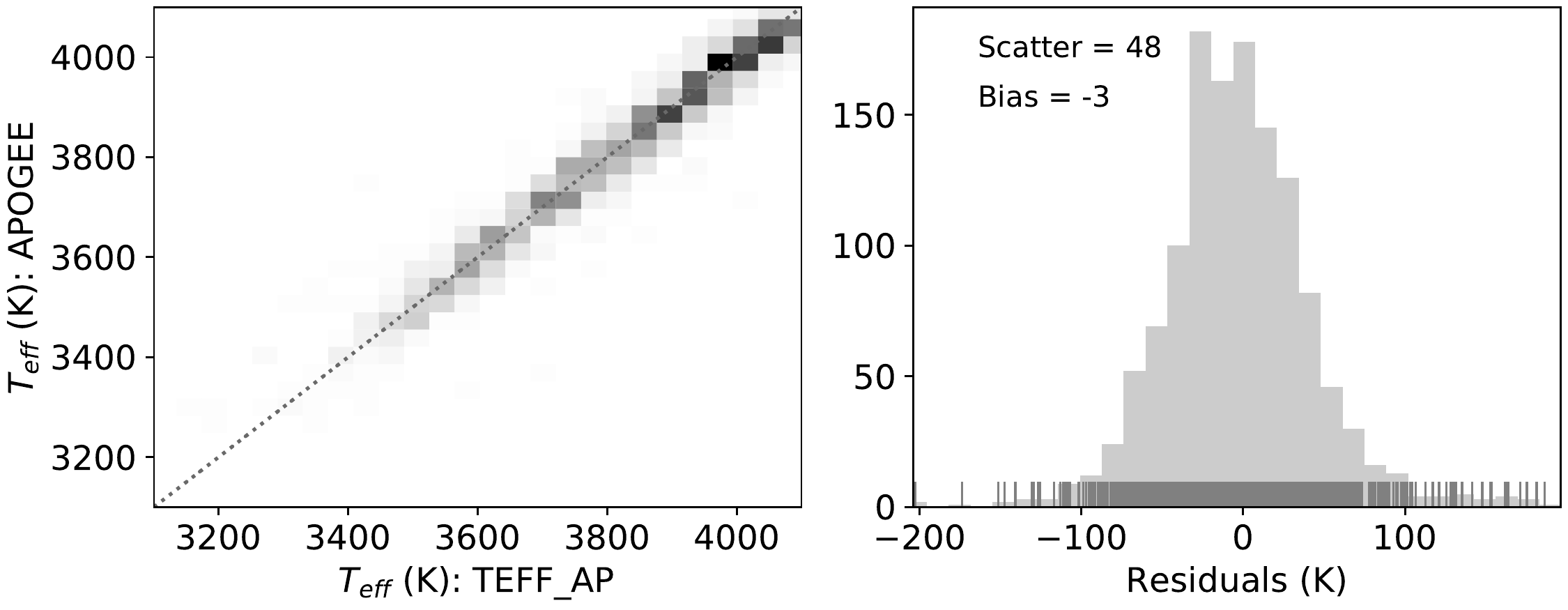}
  \plotone{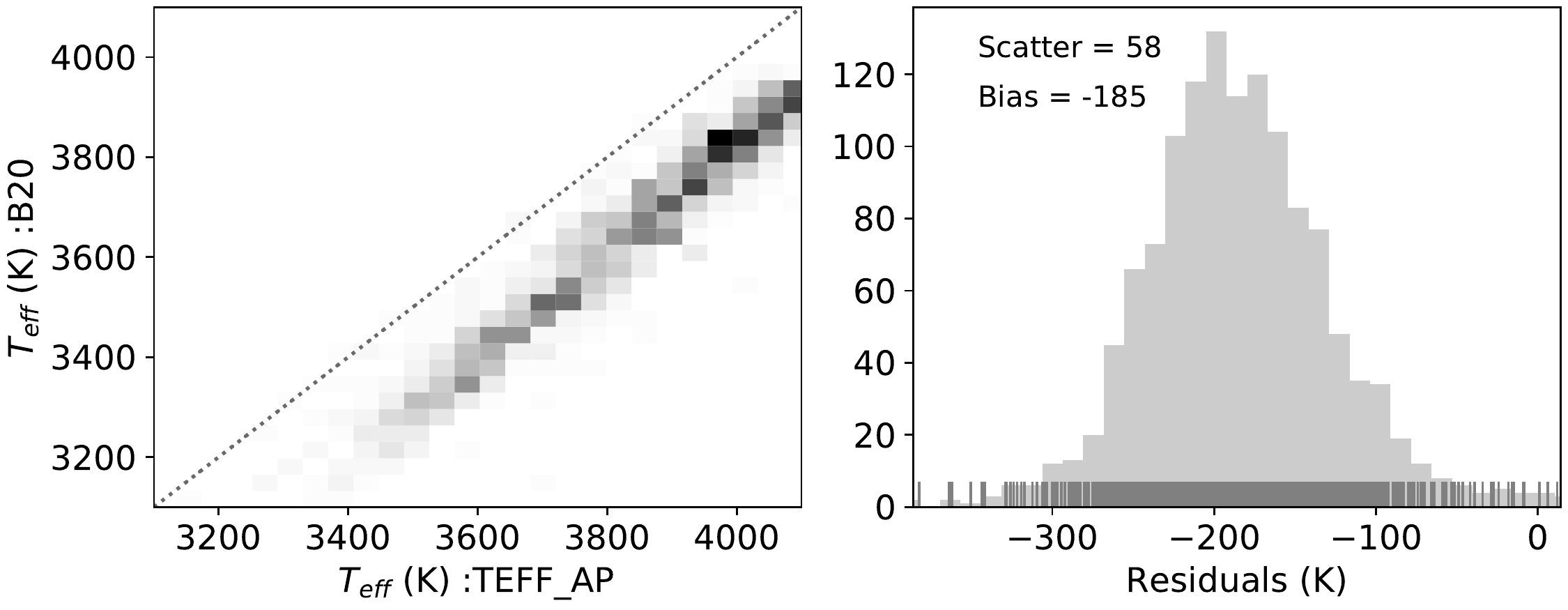}
  \plotone{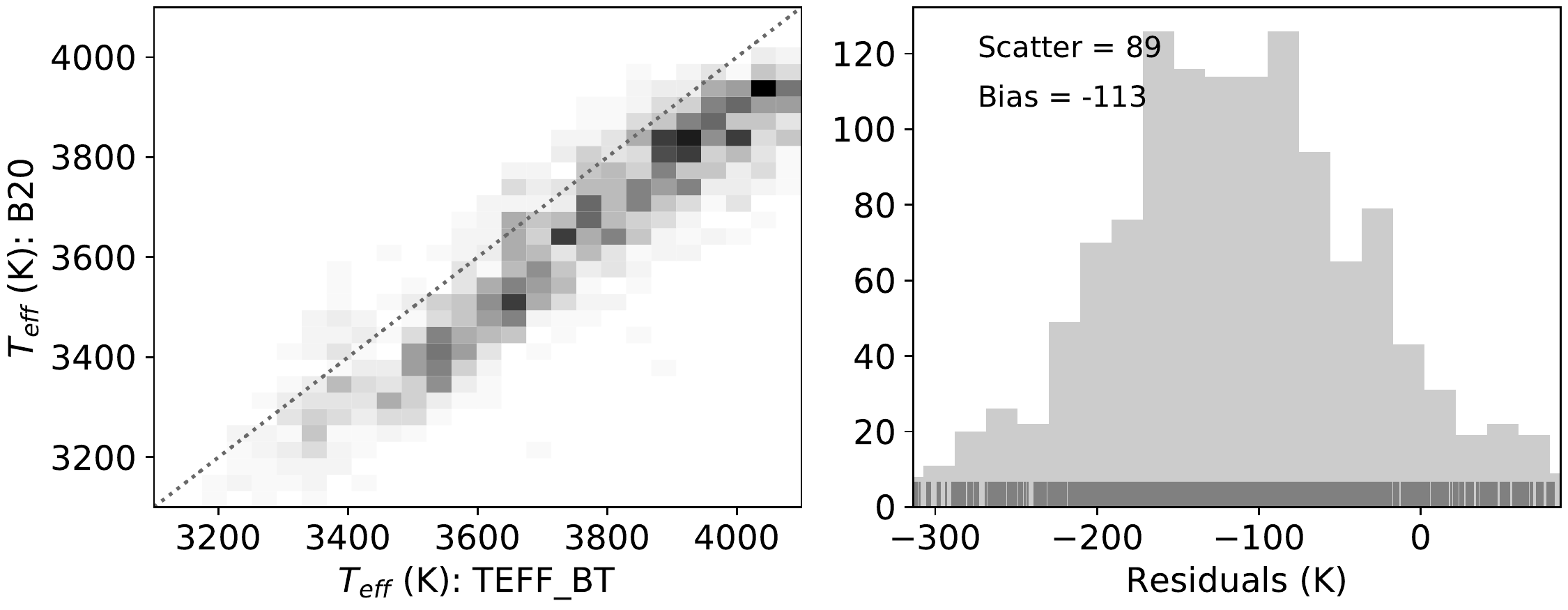}
  \caption{The top panels are comparisons of the ASPCAP-trained $T_{eff}$ (TEFF\_AP) and APOGEE stelalr labels. The top-right panel shows the distribution of their residuals. While the middel panels show the comparison between TEFF\_AP and B20.  The bottom panels show the similar comparison, but between BT-Settl-trained $T_{eff}$ (TEFF\_BT) and B20.}
  \label{fig:teff}
\end{figure*}

\begin{figure*}[hbt!]
    \centering
    \plotone{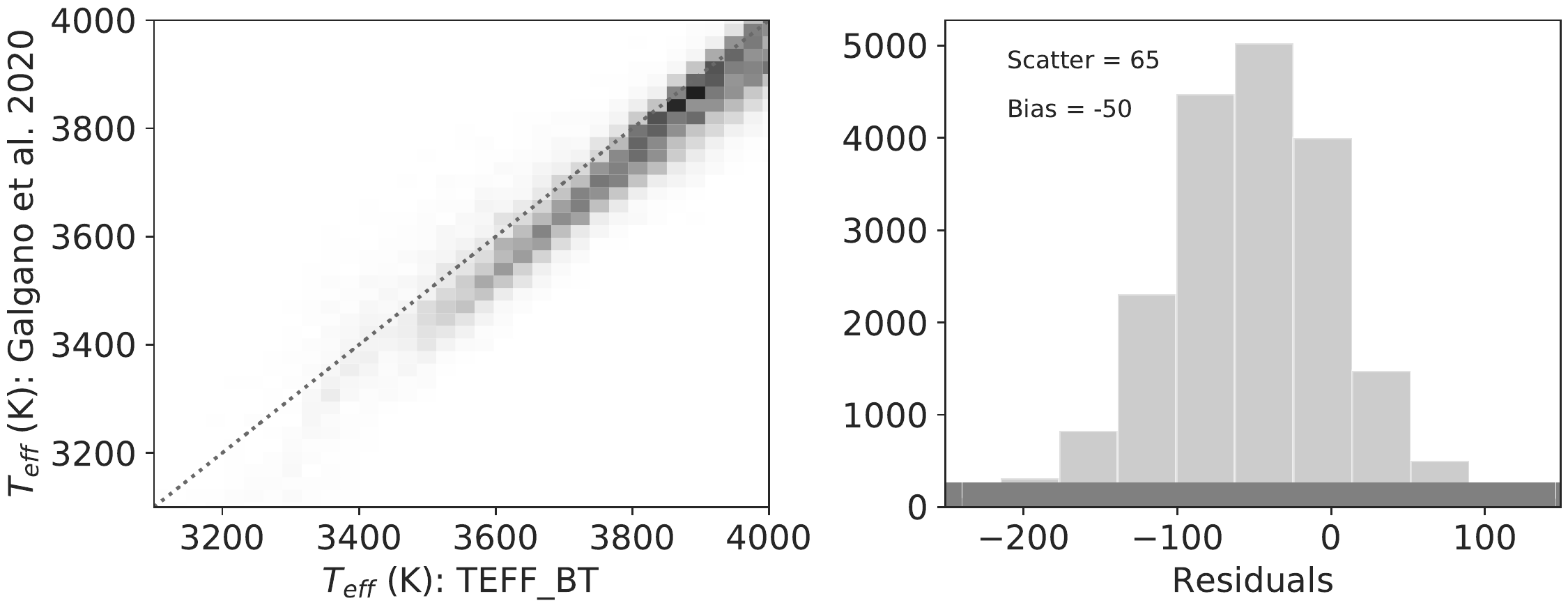}
    \caption{The left panel show the comparison between BT-Settl-trained $T_{eff}$ (TEFF\_BT) and \cite{2020AJ....159..193G}. The right panel displays the distribution of the residuals.}
    \label{fig:teffgal}
  \end{figure*}

\begin{table*}[hbt!]
  \caption{Notation of the names of the stellar parameter of LAMOST M dwarfs from different training dataset. The LAMOST spectra with APOGEE labels are used to estimate both the effective temperature and metallicities. BT-Settl synthesis spectra are used to only determine $T_{eff}$.}
  \centering
   \begin{tabular}{c c c c} 
   \hline
   Training dataset & LAMOST spectra & BT-Settl \\
    & with ASPCAP labels & synthesis spectra \\
   \hline
   effective temperature & TEFF\_AP & TEFF\_BT & \\ 
   metallicity  & M\_H\_AP & -   \\
   \hline
   \end{tabular}
  \label{table:names}
\end{table*}

Note that APOGEE stellar parameters and BT-Settl depend on different atmospheric models and are not necessarily consistent with each other. Therefore, we use them as the training data independently and predict two sets of stellar labels based on the two dataset.

We use observed (LAMOST spectra with APOGEE labels) and synthesis spectra (BT-Settl), respectively, as the training dataset. The notation of stellar parameters ($T_{eff}$, [M/H]) of LAMOST M dwarfs are named by the different training datasets as shown in Table \ref{table:names}. 

Firstly, we estimate the $T_{eff}$ and [M/H] using the APOGEE labels as the training set described in subsection \ref{subsec: aspcap teff} ($T_{eff}$) and \ref{subsec: metal} ([M/H]). We further use the synthetic spectra from BT-Settl models and obtain TEFF\_BT. More details are discussed in subsection \ref{subsec: bt_teff}.

\subsection{Effective Temperature} \label{subsec: Teff}
  \subsubsection{APOGEE temperature} \label{subsec: aspcap teff}
  First of all, we combine the LAMOST spectra with the corresponding APOGEE labels as the training dataset to train the SLAM model. Effective temperature (TEFF\_AP) and metallicity (M\_H\_AP) are determined for the test dataset by applying this SLAM model. In this section, we discuss TEFF\_AP and leave M\_H\_AP in subsection \ref{subsec: metal}.
  
  A 10-fold cross-validation (CV) is taken to estimate the precision and accuracy of the ASPCAP-trained SLAM labels. The CV-scatter and CV-bias denote the standard deviation and mean deviation respectively, and can be written as
  \begin{equation} \label{CVbias}
    {\rm CV-bias} = \frac{1}{n} \sum_{i=1}^{n}(\theta_{i,SLAM} - \theta_{i}),
  \end{equation}
  and
  \begin{equation} \label{CVscatter}
    {\rm CV-scatter} = \frac{1}{n} \sqrt{\sum_{i=1}^{n}(\theta_{i,SLAM} - \theta_{i})^2},
  \end{equation}

  where $\theta_{i,SLAM}$ is the stellar label of the $i$th star predicted by SLAM. $\theta_{i}$ denotes the stellar label of the $i$th star as ground truth. Theoretically, a robust data-driven algorithm has a small CV bias and CV scatter. The CV results displayed in Fig. \ref{fig:cv_error} indicate that the TEFF\_AP reaches a precision of 50K with no bias when SNR$_i >100$. 
  
  B20 derives spectroscopic temperatures and metallicities for 5,875 M dwarfs from the APOGEE survey. We cross-match the LAMOST data with their results and obtained 1,913 common stars. Among them, B20, LAMOST, and APOGEE DR16 together have 1,286 common stars. We compare the stellar parameters in the two catalogs separately with those we obtained using SLAM. As illustrated in the top-panel of Fig. \ref{fig:teff},  $T_{eff}$ of APOGEE DR16 and TEFF\_AP are in good agreement, which is not surprising because TEFF\_AP are estimated by stellar labels of APOGEE. The $\sim$50K scatter and 3K bias are identical to the test results of CV, which are reasonable. The mid-panel of Fig. \ref{fig:teff} shows the comparison between TEFF\_AP and B20, the residuals between the two have a dispersion of $\sim 60$K, and an offset of 185K. This bias is mainly due to the 182K difference between the temperature of APOGEE and B20. This is the result of different stellar atmospheric models used in the stellar parametrization.

  \subsubsection{BT-Settl temperature} \label{subsec: bt_teff}

  We then set up the alternative training dataset using BT-Settl synthetic spectra so that the lower $T_{eff}$ can go down beyond 3000K. We follow the preprocessing procedure developed by \cite{2020ApJS..246....9Z}\footnote{\url{https://github.com/hypergravity/astroslam}} to adjust the resolution and wavelength to be same as LAMOST low-resolution spectra. The training labels of the model grids\footnote{\url{https://phoenix.ens-lyon.fr/Grids/BT-Settl/CIFIST2011b}} are $2200<{\rm T_{eff}}<7000$K, $-1.0<{\rm [M/H]}<0.0$ dex, $2.5<{\rm \log g}<5.5$ dex with steps of 100K, 0.5 dex and 0.5 dex, respectively. 

  The original grid from BT-Settl is too sparse. So we firstly interpolate the grid to obtain a new training dataset with denser grids. Because SLAM is a forward model, it can be used to do the interpolation. We randomly draw 15,000 points in the parameter space with $2800<{\rm T_{eff}}<4500$ K, $-1.0<{\rm [M/H]}<0.0$ dex and $4.5<{\rm \log g}<5.5$ dex following the uniform distributions. And the corresponding synthetic spectra are obtained from the SLAM model \textit{SLAM\_0}, which is trained by the sparse original grid of BT-Settl. We then train model \textit{SLAM\_1} with the 15,000 synthetic spectra interpolated from model \textit{SLAM\_0} as training dataset. Finally, we predict TEFF\_BT for the LAMOST spectra using \textit{SLAM\_1}.
  
  Similar to subsection \ref{subsec: aspcap teff}, 10-fold cross-validation is used to test the performance and robustness of our model. Random Gaussian noise is added to the test spectra. The left panel of Fig. \ref{fig:cv_error} shows that CV errors change with a given signal-to-noise ratio (SNR). We obtain a random error of 40K with CV-bias of 5K at SNR $> 100$. The results of the CV errors indicate that our method effectively works. Fig. \ref{fig:cv_error} also shows that CV-scatter varies with SNR. Note that the SNR is calculated by noise artificially added on fluxes of the normalized BT-Settl synthetic spectra. 
  
  We compare the temperatures with literature to assess the performance of our approach. Comparisons with B20 show that TEFF\_BT are in agreement with their results with a scatter of  $\sim$90K and an offset of $\sim$110K (see Fig. \ref{fig:teff}). \cite{2018AJ....155..180M} presented {\it TESS} Cool Dwarf Catalog providing 1,140,255 cool dwarfs with $T_{eff}$ determined on the basis of the empirical color-temperature relations of \cite{2015ApJ...804...64M}. We cross-match LAMOST M dwarfs with this cool dwarf catalog and compare the stellar parameters. We find that TEFF\_BT agrees with the {\it TESS} catalog with a scatter of 114K and an offset of $\sim $100K. This is similar to the comparison with B20, which is not surprising since the $T_{eff}$ of both B20 and {\it TESS} Cool Dwarf Catalog are calibrated with \cite{2015ApJ...804...64M}. Finally, TEFF\_BT is compared with \cite{2020AJ....159..193G}, which estimate stellar parameters of $\sim$ 30,000 M dwarfs from LAMOST DR1 based on {\it TESS} Cool Dwarf Catalog, as shown in Fig. \ref{fig:teffgal}. As expected, the residuals of $T_{eff}$ in Fig. \ref{fig:teffgal} demonstrate a small scatter of 65K and an offset of 50. Compared to other results, the smaller difference is due to the fact that both \cite{2020AJ....159..193G} and we used the same LAMSOT spectra to estimate the effective temperatures.

 \subsection{Metallicity} \label{subsec: metal}

  \begin{figure*}[hbt!]
    \plotone{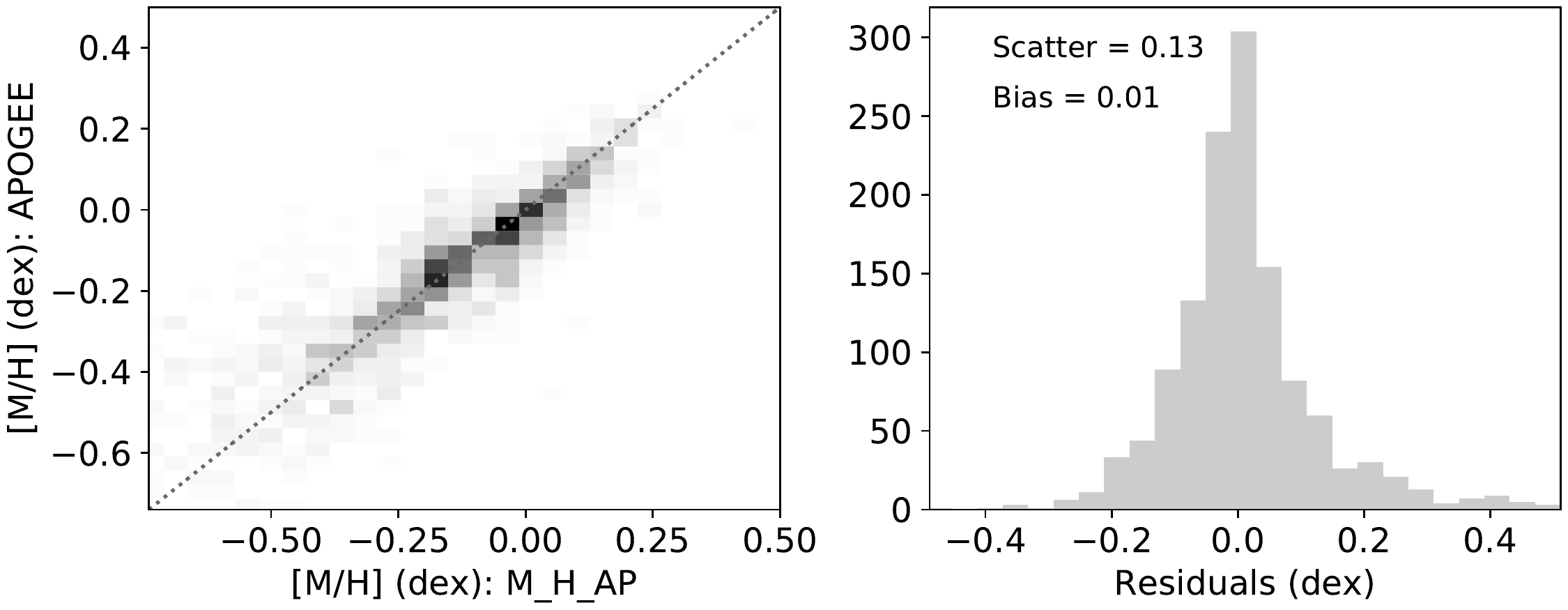}
    \plotone{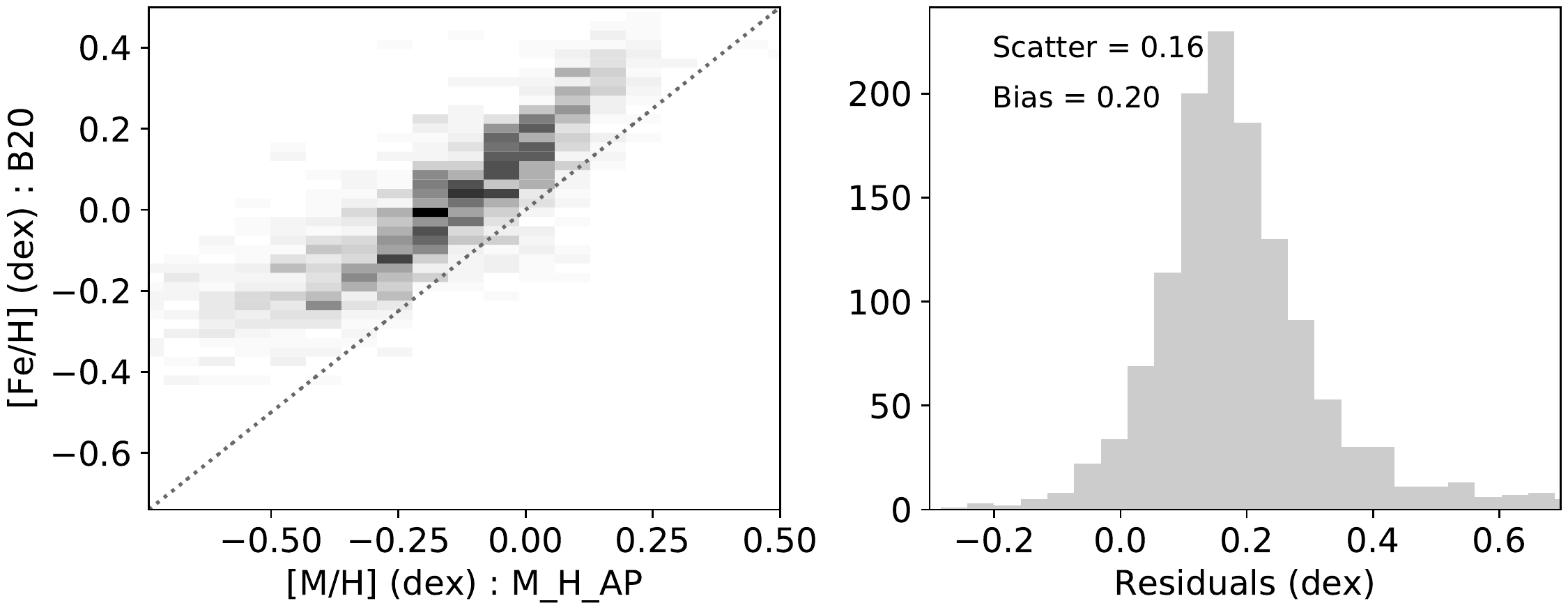}
    \plotone{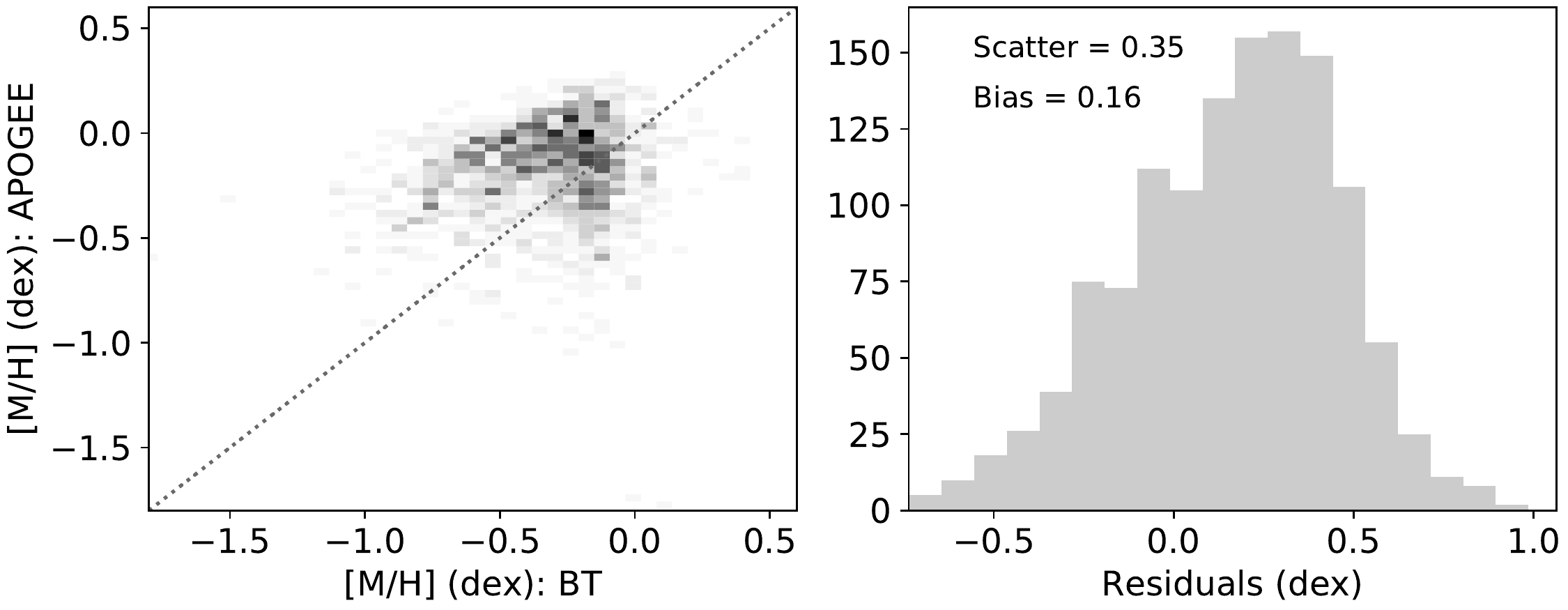}
    \caption{The top-left panel shows the comparison between APOGEE-trained [M/H] (M\_H\_AP) and metallicities of APOGEE. The top-right panel displays the residuals of the metallicities with a scatter of 0.13 dex and no bias. The middel panels display the comparison between M\_H\_AP and B20. The residuals of M\_H\_AP and B20 shows a 0.16 dex scatter and a 0.2 dex bias. The bottom panel shows the comparision with metallicities derived by BT-Settl synthetic spectra and  [M/H] of APOGEE, they have a 0.35 dex scatter and a 0.16 dex offset.
    }
    \label{fig:mh}
  \end{figure*}

\begin{figure*}[hbt!]
    \centering
    \includegraphics[height=8.25cm]{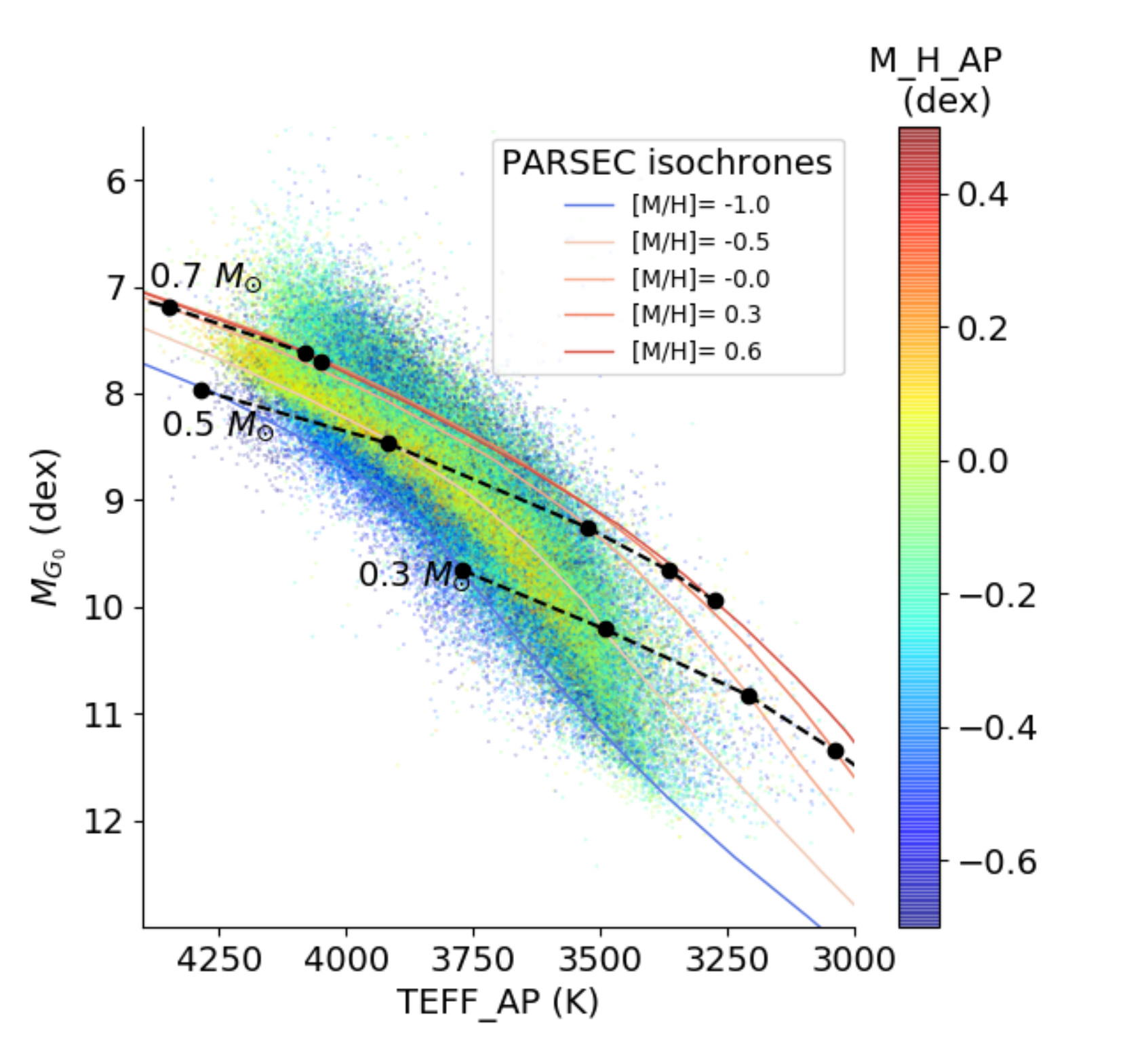}
    \includegraphics[height=8.25cm]{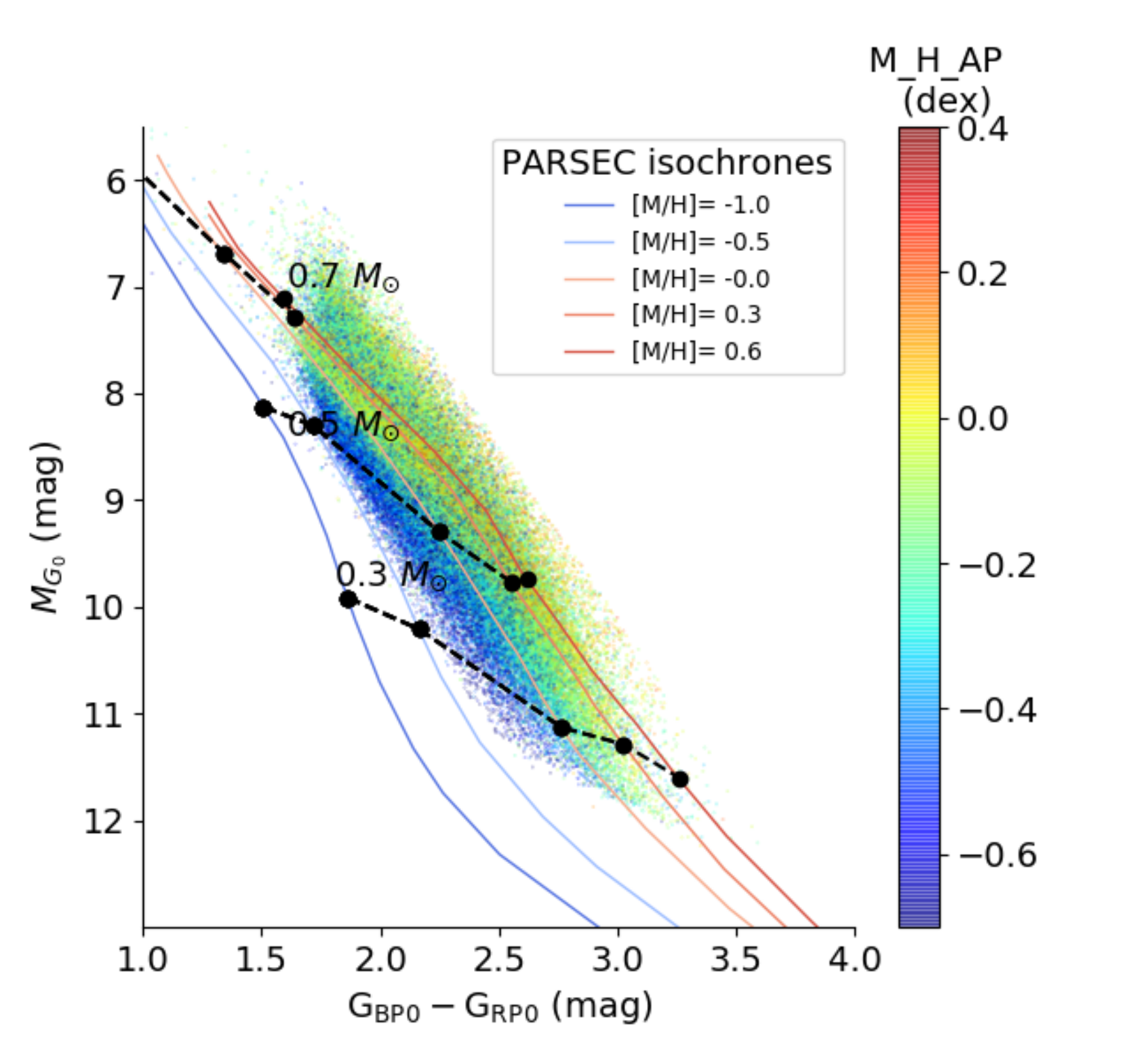}
    \includegraphics[height=8.25cm]{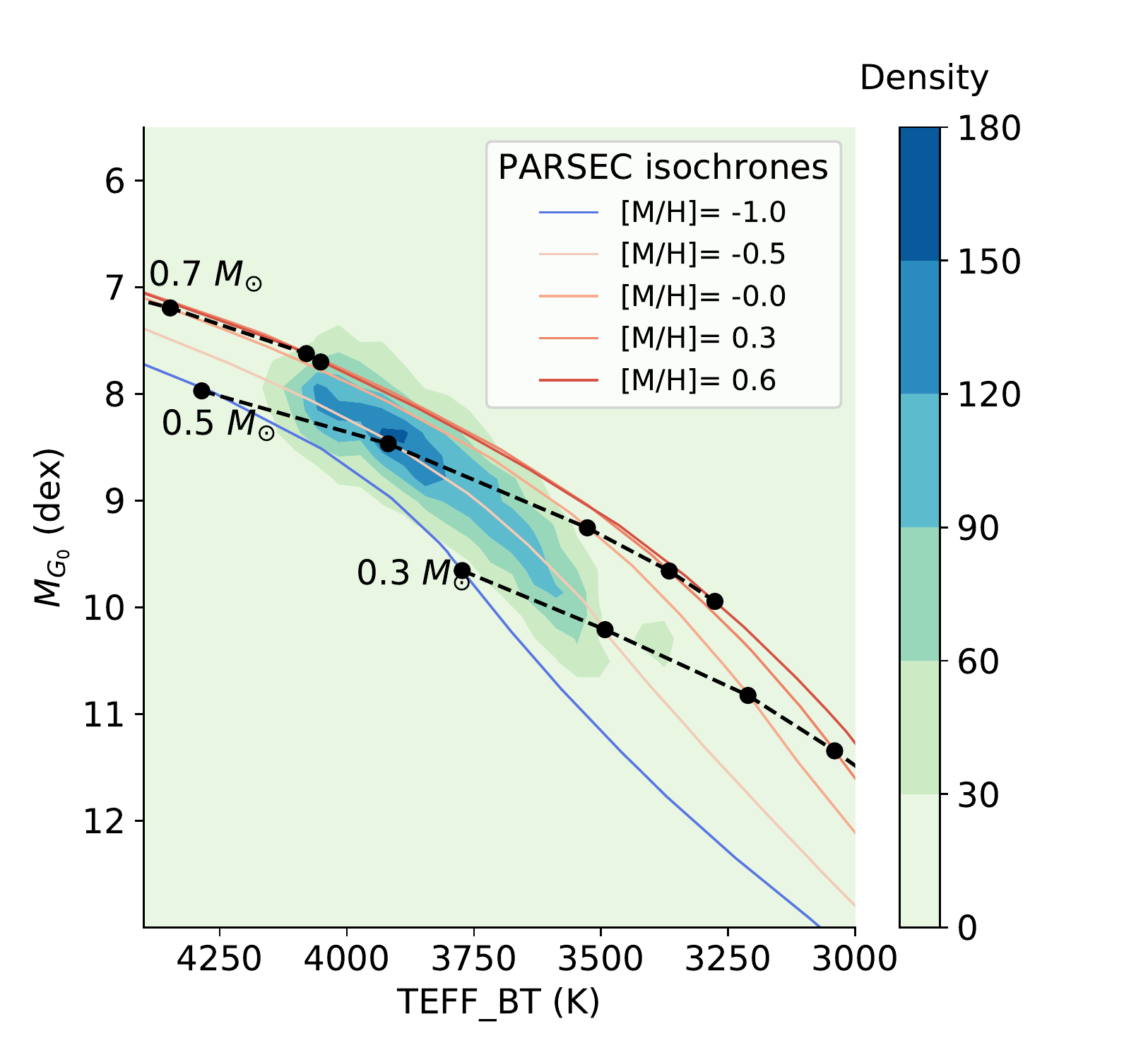}
    \caption{The top-left panel displays the HR-diagram of $\sim$ 9,000 M dwarf stars with SNR$_i > 50$, a sub-sample of our catalog in the SLAM $T_{eff}$ (TEFF\_AP) versus {\it Gaia} $M_G$ plane, and is colored by M\_H\_AP. The top-right pabel shows the same HRD as the top-left panel, but the $x$-axis is $G_{BP}-G_{RP}$ colors. The bottom panel displays the contours drawn with the same stars as the left panel, while the temperatures are TEFF\_BT. The solid lines indicate the PARSEC isochrones from [M/H]=-1.0 dex to [M/H]=0.6 dex. The grey dashed curves represent the locations of 0.3, 0.5 and 0.7 $M_{\sun}$. 
    }
    \label{fig:hrd_bt_kde}
 \end{figure*}
 
  Before estimating the metallicities of LAMOST M dwarfs, we first compared the [M/H] of APOGEE DR16 and [Fe/H] of B20, with a residual difference of only 0.1 dex, while B20 is overall more metal-rich than APOGEE  by 0.2 dex. We find metallicity of APOGEE and B20 have better consistency where the metallicity abundance is higher. For stars with [M/H] larger than -0.25, the scatter between B20 and APOGEE is 0.07 dex with an offset of 0.15 dex. While for stars with [M/H] $< -0.25$,  the scatter and bias are 0.15 and 0.25 dex, respectively.
  
  M\_H\_AP is determined by [M/H] of APOGEE as the training dataset. The prediction of M\_H\_AP is estimated at the same time as the TEFF\_AP using SLAM. We find M\_H\_AP agrees well with APOGEE labels (see the top-panel of Fig. \ref{fig:mh}). Their good agreement is exactly what we expected, since the APOGEE parameters are used as the training labels. Furthermore, M\_H\_AP is in agreement with B20 metallicity with a scatter of 0.16 dex as shown in the mid-panel of Fig. \ref{fig:mh}. We also find a similar scenario to the comparison between APOGEE and B20. For stars with [M/H] $> -0.25$, the scatter is $\sim$0.1 dex with a bias of 0.14 dex, but for stars with metallicities lower than -0.25 dex , the scatter is 0.2 dex with a 0.3 dex offset.

  Metallicities could be also estimated by the BT-Settl models together with $T_{eff}$. However, the derived BT-Settl [M/H] (hereafter, M\_H\_BT) shows no clear correlation with observed data published in previous studies (see the bottom-panel of Fig \ref{fig:mh}). In the metal-poor regime ([M/H] $<-0.25$ dex), though with larger scatter, the correlation between M\_H\_BT and APOGEE metallicities is obvious. But for [M/H] $> -0.25$ dex, M\_H\_BT is not able to distinguish well the metallicities of stars. This is probably due to the lack of model grids with [M/H] $>0$. Therefore, the BT-Settl-trained [M/H] is not adopted in our catalog Table \ref{tab:all}.


\subsection{Hertzsprung-Russell Diagram} \label{subsec: hrd}

  Fig. \ref{fig:hrd_bt_kde} displays the Hertzsprung-Russell Diagram (HRD) of the LAMOST M dwarfs in the \textit{Gaia} $M_G$ versus logarithmic $T_{eff}$ (TEFF\_AP) plane. Note that the $M_G$ is estimated from the Bayesian distance from \cite{Bailer-Jones2018} with extinction removed. 3D dust-reddening maps from \textit{Bayestar} \citep{2019ApJ...887...93G} is used to estimate the visual extinction $A_V$ for each star. $A_G$ is further estimated from $A_V$ using the extinction factor from \cite{2019ApJ...877..116W}. Fig. \ref{fig:hrd_bt_kde} is the de-reddened HRD for a sub-sample of $\sim 9,000$ M dwarf stars in our catalog. As illustrated in the top-left panel, the metallicity (M\_H\_AP) can be clearly distinguished in HRD. Moreover, the top-right panel shows the color-magnitude diagram in $G_{BP}-G_{RP}$ versus $M_G$ plane, the gradient of metallicity is still very clear and shows the similar trend as in the top-left panel.
  
  We further overlap the PARSEC (the Padova and Trieste Stellar Evolutionary Code) \citep{2012MNRAS.427..127B} theoretical tracks with the age of 1 Gyr and find that they are well fit with each other as shown in the bottom panel of Fig. \ref{fig:hrd_bt_kde}. The PARSEC version 1.2S\footnote{\url{http://stev.oapd.inaf.it/cgi-bin/cmd}} \citep{2014MNRAS.444.2525C} provides revisions on very-low-mass stars (VLMs) from the BT-Settl model with a wide range of metallicities from -2.19 to +0.70 dex. As displayed in Fig. \ref{fig:hrd_bt_kde}, we found: a) The HRD given by the PARSEC stellar model shows good agreement with our observed HRD for TEFF\_AP and TEFF\_BT; b) Stars above the isochrone of [M/H] = 0.6 are likely to be in binary systems. 
  
\subsection{Chromosperic Activity} \label{subsec: activity}
Considering active M dwarf is the only class of stars for which magnetic field affects overall stellar parameters systematically \cite{2021A&ARv..29....1K}. H$\alpha$ emission which can be the indicator of chromospheric activity might alter the reliability of the stellar parametrization. \cite{2015RAA....15.1182G} found that the fraction of active stars increases as spectral subtype becomes later. In our work, there is $\sim 8$ percent of M dwarf stars that have active magnetic fields. So during the procedure of data pre-processing, the pixels at the wavelength of H$\alpha$ emission are masked to improve the precision of stellar parameter estimation.

\subsection{The Catalog}
\begin{table}[hb]
  \caption{The field definition of the stellar parameters catalog of LAMOST M dwarf stars.}
  \centering
  \begin{tabular}{lcc}
  \hline
  \multicolumn{1}{c}{Column} & Unit           & \multicolumn{1}{c}{Description}                         \\ \hline
  source\_id                 &                & \textit{Gaia} identification ID                        \\
  obsid                      &      & LAMOST unique spectra ID                                       \\
  ra\_obs                    & deg  & LAMOST fiber pointing right ascension                   \\
  dec\_obs                   & deg  & LAMOST fiber pointing declination                       \\
  snru                       &            & LAMOST signal noise at SDSS $u$-band                  \\
  snrg                       &            & LAMOST signal noise at SDSS $g$-band                   \\
  snrr                       &                & LAMOST signal noise at SDSS $r$-band                   \\
  snri                       &                & LAMOST signal noise at SDSS $i$-band                   \\
  snrz                       &                &LAMOST signal noise at SDSS $z$-band                   \\
  z                          &                & LAMOST redshift                                         \\
  z\_err                     &                & LAMOST redshift uncertainty                             \\
  type                       &                & magnetic activity                                       \\
  TEFF\_BT                   & K    & effective temperature from BT-Settl-trained SLAM        \\
  TEFF\_BT\_ERR              & K    & uncertainty of effective temperature from BT-Settl-trained SLAM \\
  TEFF\_AP                   & K    & effective temperature from ASPCAP-trained SLAM        \\
  TEFF\_AP\_ERR              & K    & uncertainty of effective temperature from ASPCAP-trained SLAM   \\
  M\_H\_AP                   & dex  & {[}M/H{]} from ASPCAP-trained SLAM                      \\
  M\_H\_AP\_ERR              & dex  & uncertainty of {[}M/H{]} from ASPCAP-trained SLAM     \\
  \hline
  \end{tabular}
  \label{tab:catalog-header}
\end{table}

Table \ref{tab:catalog-header} shows the field definition of the stellar parameter catalog of our results. The important information of both the LAMOST and \textit{Gaia} observations are presented in our catalog. TEFF\_AP, TEFF\_AP\_ERR, M\_H\_AP and M\_H\_ERR are the ASPCAP-trained SLAM effective temperature and metallicity with the corresponding uncertainty. TEFF\_BT and TEFF\_BT\_ERR are the BT-Settl-trained SLAM $T_{eff}$ with the estimated uncertainty. Note that the \textit{type} column is adopted from \citet{2015RAA....15.1182G}, which indicates the magnetic activity determined by measuring H$\alpha$ activity. The stellar parameters including $T_{eff}$ and [M/H] with the corresponding uncertainties of LAMOST M dwarfs estimated in this work are presented in Table \ref{tab:all}.


\section{Discussion and conclusions} \label{sec:discussion}

\subsection{The accuracy of metallicity assessed from Open Clusters}

\begin{figure}[hbt!]
    \centering
    \includegraphics[width=10cm]{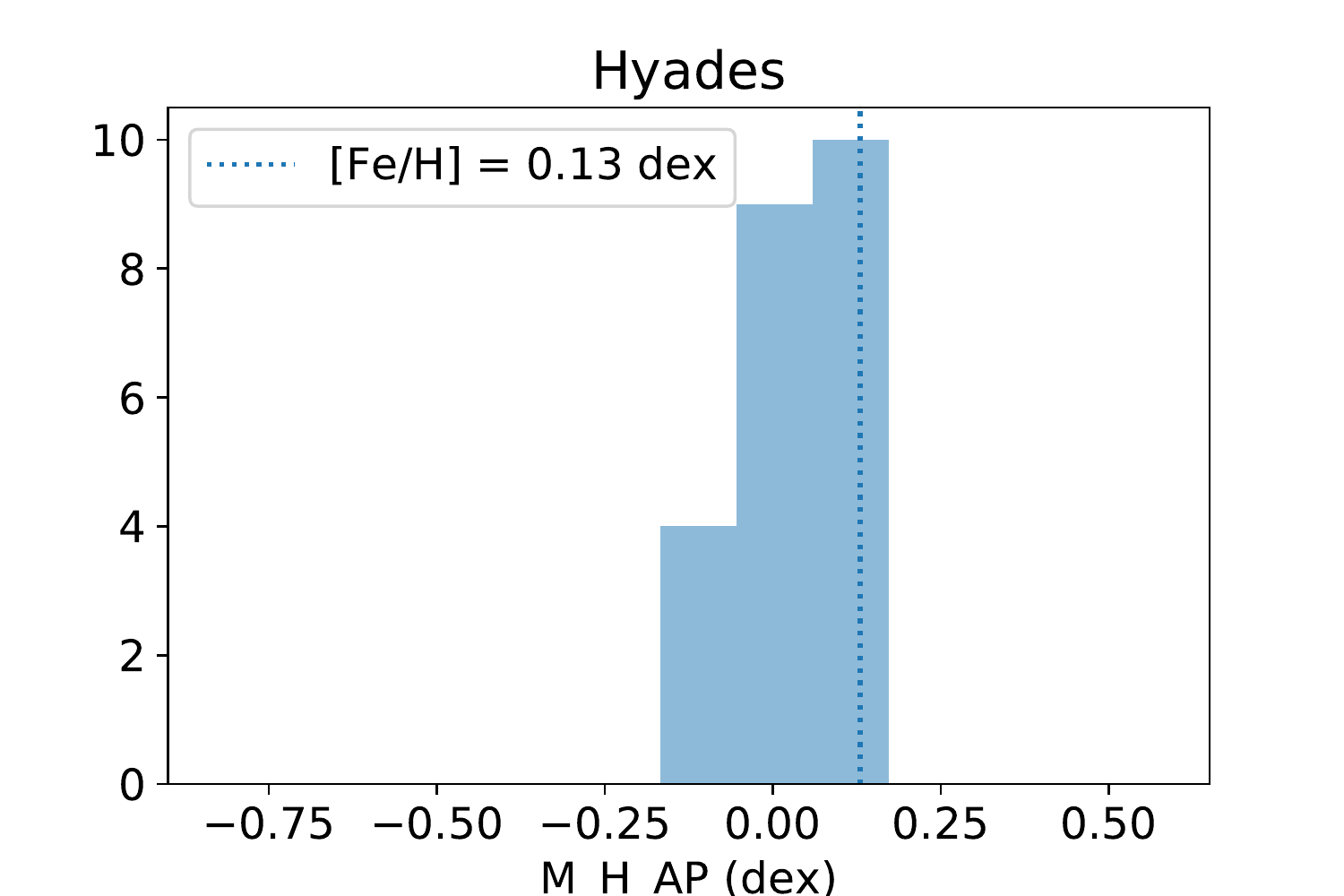}
    \caption{Metallicity distribution in Hyades. The vertical line indicate the metallicity [Fe/H]=0.13 given by previous studies. 
    }
    \label{fig:hyades}
 \end{figure}
 
 \begin{figure*}[hbt!]
    \plotone{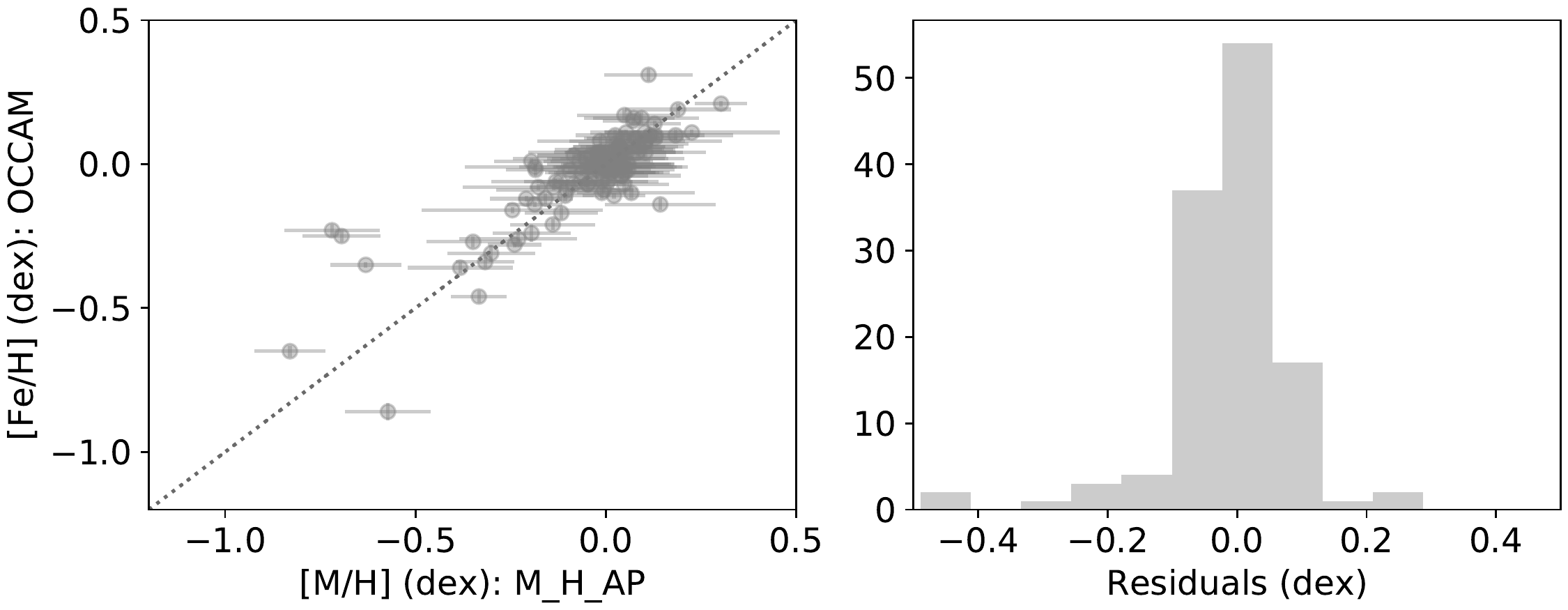}
    \caption{The left panel shows the comparison of M\_H\_AP and [Fe/H] determined by \cite{2020AJ....159..199D}. The right panel displays the distribution of [M/H] difference.
    }
    \label{fig:opencluster}
 \end{figure*}
 
 \begin{table}[hb]
  \caption{The comparison of [M/H] of LAMOST M dwarf stars with OCCAM. $<$M\_H\_AP$>$ is the mean metallicity in each cluster derived by our work, ${\rm [Fe/H]}_{\rm OCCAM}$ is the mean value given by OCCAM and $N$ is the number of the member in the cluster.}
  \centering
 \begin{tabular}{lccrr}
\toprule
    Cluster &  $<$M\_H\_AP$>$ &  ${\rm [Fe/H]}_{\rm OCCAM}$ &   $N$ \\
\hline
ASCC 16       &   -0.07 $\pm$ 0.03 &    -0.04 $\pm$ 0.03 &            5 \\
ASCC 21       &   -0.21  $\pm$ 0.03 &   -0.12 $\pm$ 0.04 &            2 \\
Berkeley 19   &   -0.19  &            -0.01 &              1 \\
Berkeley 29   &    0.08  &            0.09 &              1 \\
Briceno 1     &   -0.06  &        -0.03 &              1 \\
Chupina 1     &   -0.63  &       -0.35 &               1 \\
Chupina 3     &   -0.11  &       -0.11 &             1 \\
Chupina 5     &   -0.83  &       -0.65 &              1 \\
Collinder 69  &   -0.20  $\pm$ 0.08 &  -0.21 $\pm$  0.14&      5 \\
Collinder 70  &    0.01  &       -0.01 &            1 \\
Koposov 62    &   -0.05  &        0.07 &            1 \\
Melotte 20    &    0.05 $\pm$ 0.11 &    0.05 $\pm$  0.12 &       21 \\
Melotte 22    &    0.00 $\pm$ 0.13 &    0.00 $\pm$  0.09&        66 \\
NGC 2420      &   -0.23 &         -0.26 &              1 \\
NGC 2682      &   -0.16 $\pm$ 0.25 &    -0.22 $\pm$ 0.30 &     7 \\
NGC 752       &   -0.17 $\pm$ 0.26 &    -0.09 $\pm$ 0.11 &     6 \\
\hline
\end{tabular}
\label{tab:opcl}
 \end{table}
 
To assess the accuracy of the metallicities estimated in our work, we select 23 Hyades member stars from our catalog. Hyades is the nearest open cluster, as far as we know \citep{2009A&A...497..209V}, with a metallicity of about +0.13 dex \citep{2006AJ....131.1057S}. We select all stars in a circle with a radius of 5 degrees, centered on right ascension of 66.725$^\circ$ and declination of 15.867$^\circ$ from {\it Gaia} DR2. Then, we adopt that stars located within $4<${\tt\string pmra / parallax}$<6$, $-2.5<${\tt\string pmdec / parallax}$<0$ and $0.04<${\tt\string parallax}$<0.05$, where {\tt\string pmra} and {\tt\string pmdec} are {\it Gaia} proper motions, are the member stars of Hyades. We cross-match our catalog with this sample and obtain 23 M dwarf member stars. We find that the distribution of M\_H\_AP of the members has a mean of 0.0 dex and a standard deviation of 0.1 dex, as shown in Fig. \ref{fig:hyades}. This result tentatively verifies that the accuracy of M\_H\_AP is around 0.1 dex. 
 
Furthermore, we cross-match our catalog with the Open Cluster Chemical Analysis and Mapping (OCCAM) survey \citep{2020AJ....159..199D} and obtain 138 member stars belonging to 15 open clusters. Fig. \ref{fig:opencluster} compares [Fe/H] given by \cite{2020AJ....159..199D} with M\_H\_AP. The left panel shows that M\_H\_AP matches very well with the metallicity of the corresponding clusters in the ranges from -0.7 to 0.4\,dex. A few stars with higher metallicity in literature are estimated as lower metallicity using our method. This may implies some limit of the estimation in metal-poor regime. The right panel of Fig. \ref{fig:opencluster} displays the distribution of residuals between M\_H\_AP and [Fe/H] of OCCAM. The mean value is 0.01 dex and the standard deviation is 0.1 dex. This illustrates that the uncertainty of the metallicity derived in this work is around 0.1 dex. Detailed information on the comparison of metallicities of cluster members is shown in Table. \ref{tab:opcl}.
 
\subsection{Precision of metallicity from wide binary}
  \begin{figure*}[hbt!]
    \plotone{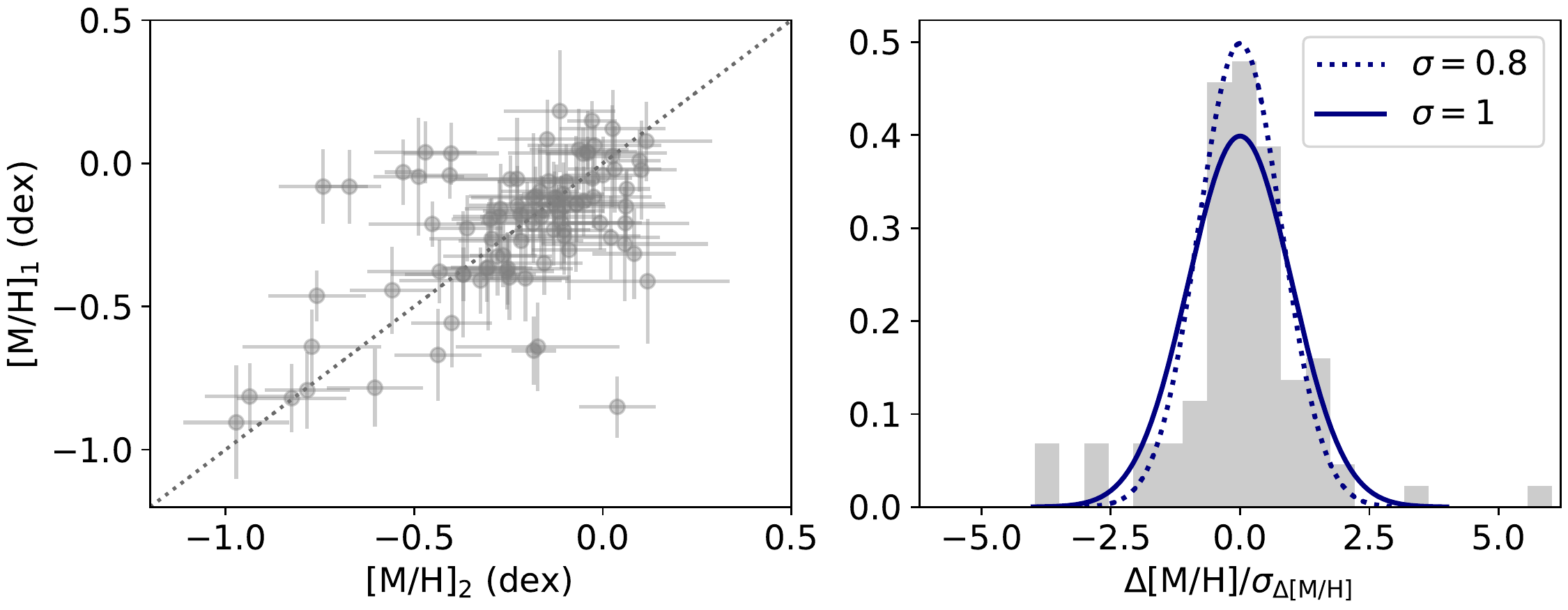}
    \caption{Comparison of metallicities of high-confidence binaries with separations less than 3,000 au. Left: comparison of the metallicities (M\_H\_AP) of the primary and secondary component. Right: distributions of uncertainty-normalized [M/H] difference, compared to Gaussians with the corresponding $\sigma$.}
    \label{fig:wide binary}
 \end{figure*}
 
 We further assess the precision of the metallicity using M dwarf - M dwarf wide binaries (Qiu et al., in prep). We start with the catalog of initial wide binaries candidates released by \cite{2020ApJS..246....4T}, which contains 807,611 candidates, selected form {\it Gaia} DR2 within a distance of 4.0 kpc and a maximum projected separation $s=1.0$ pc. This catalog contains many types of wide binaries (e.g., main sequence - main sequence, main sequence - white dwarf, white dwarf - white dwarf etc.), and these wide binary stars may be polluted by visual binaries (chance alignments) at large separations, as described in Section 3.5 in \cite{2020ApJS..246....4T}. We finally choose 92 pairs of binary stars with the separations less than 3,000 au to assess the precision of [M/H].

 Fig. \ref{fig:wide binary} compares the differences of the metallicities (M\_H\_AP) of the companions of these binaries. M\_H\_AP is expected to be consistent with each other if the companions are physically associated. The left panel shows the pairs have similar [M/H], only 2 of the 92 systems fall outside of 3-sigma range. These outliers are most likely not physical binary stars. The mean difference of metallicity of the primary and secondary companions is 0.01 dex with scatter of 0.23 dex. The right panel is the uncertainty-normalized metallicity difference; i.e., $\Delta {\rm [M/H]} / \sigma_{\Delta {\rm [M/H]}} = ({\rm [M/H]_1} - {\rm [M/H]_2})/\sqrt{\sigma_{\rm [M/H]_1^2} + \sigma_{\rm [M/H]_2^2}}$. If the derived $\sigma_{\rm [M/H]}$ values are accurate, the uncertainty-normalized metallicity difference should be distributed as a Gaussian distribution with $\sigma =1$. As shown in the right panel, the distribution matches better with $\sigma \sim 0.8$, which suggests that M\_H\_AP uncertainties may be overestimated by $\sim 20 \%$.

\subsection{Summary} \label{subsec: summary}
In this work, we have derived a spectroscopic catalog of stellar parameters for $\sim 300,000$ M dwarf stars. Precise effective temperatures and metallicities of M dwarf stars from LAMOST DR6 and {\it Gaia} DR2 with are given in this catalog. Stars located within the range of $2800<T_{eff}<4500$K (both of TEFF\_AP and TEFF\_BT) are finally adopted in our catalog. Two versions of effective temperatures are obtained with precisions of 40K for TEFF\_BT and 50K for TEFF\_AP at SNR $>50$. Particularly, TEFF\_AP agrees with B20 with a 60K scatter and a 185K offset. The systematic errors come from different stellar atmospheric models, in this case, BT-Settl model and MARCS, respectively.

This study provides a method using BT-Settl model to obtain the parameters of LAMOST M dwarf stars. We also publish the code and the stellar parametrization pipeline on the website\footnote{\url{https://github.com/jiadonglee/MDwarfMachine}}. Note that the data-driven method to derive stellar labels strongly relies on training datasets, for the estimation in this work is on BT-Settl model and stellar parameters of APOGEE. We address that SLAM can be used as the industrial framework against which to decode the stellar parameters of the cool atmosphere of M dwarf stars in the upcoming surveys such as SDSS-V \citep{2017arXiv171103234K}.

 \acknowledgments
  We thank the anonymous referee for the very hepful comments. This work is supported by National Key R\&D Program of China No. 2019YFA0405500. C.L. thanks the National Natural Science Foundation of China (NSFC) with grant No. 11835057. Guoshoujing Telescope (the Large Sky Area Multi-Object Fiber Spectroscopic Telescope LAMOST) is a National Major Scientific Project built by the Chinese Academy of Sciences. Funding for the project has been provided by the National Development and Reform Commission. LAMOST is operated and managed by the National Astronomical Observatories, Chinese Academy of Sciences. This work has made use of data from the European Space Agency (ESA) mission {\it Gaia} (\url{https://www.cosmos.esa.int/gaia}), processed by the {\it Gaia} Data Processing and Analysis Consortium (DPAC, \url{https://www.cosmos.esa.int/web/gaia/dpac/consortium}). Funding for the DPAC has been provided by national institutions, in particular the institutions participating in the {\it Gaia} Multilateral Agreement. 
  
  %
  
  \vspace{5mm}
  \facilities{LAMOST, \textit{Gaia}}
  
  
  \software{astropy \citep{2018AJ....156..123A}, scipy \citep{2019arXiv190710121V}, scikit-learn \citep{scikit-learn}, SLAM \citep{2020ApJS..246....9Z}, TOPCAT \citep{2005ASPC..347...29T}}



\appendix
\section{catalog}

\begin{table}[]
\caption{The stellar parameters ($T_{eff}$, [M/H]) of the LAMOST M dwarfs catalog. The complete table can be found online \url{http://paperdata.china-vo.org/jordan/Mdwarf/jdli21.fits}.}
    \centering
    \begin{tabular}{lcccccccc}
\toprule
    obsid & TEFF\_BT & TEFF\_BT\_ERR & TEFF\_AP & TEFF\_AP\_ERR & M\_H\_AP & M\_H\_AP\_ERR & snri \\
      \hline
300208242 &    3599 &          38 &    3804 &          66 &   0.14 &       0.16 &   46 \\
331715172 &    3676 &          38 &    3746 &          73 &  -0.17 &       0.17 &   39 \\
438102143 &    3587 &          36 &    3602 &          51 &  -0.21 &       0.12 &   74 \\
285016161 &    3628 &          38 &    3612 &          76 &  -1.12 &       0.18 &   36 \\
185605033 &    3680 &          38 &    3763 &          77 &  -0.23 &       0.18 &   35 \\
400108239 &    3681 &          36 &    3749 &          49 &  -0.15 &       0.12 &   78 \\
  8003208 &    3584 &          38 &    3704 &          72 &  -0.25 &       0.17 &   40 \\
364813026 &    3645 &          38 &    3777 &          85 &   0.03 &       0.20 &   29 \\
387207221 &    3522 &          36 &    3612 &          47 &  -0.14 &       0.12 &   86 \\
574612100 &    3451 &          38 &    3425 &          97 &  -0.52 &       0.22 &   23 \\
 21803121 &    3602 &          38 &    3692 &          76 &  -0.14 &       0.18 &   36 \\
196704044 &    3606 &          38 &    3680 &          59 &  -0.20 &       0.14 &   57 \\
212806086 &    3545 &          36 &    3669 &          48 &   0.09 &       0.12 &   82 \\
593209154 &    3635 &          36 &    3724 &          45 &  -0.08 &       0.11 &   92 \\
600415191 &    3645 &          36 &    3726 &          35 &  -0.18 &       0.09 &  144 \\
333012162 &    3280 &          38 &    3472 &          92 &  -0.18 &       0.21 &   25 \\
287010071 &    3380 &          38 &    3649 &          81 &   0.06 &       0.19 &   32 \\
334502112 &    3643 &          38 &    3677 &          67 &  -0.38 &       0.16 &   45 \\
353612149 &    3384 &          38 &    3474 &          63 &  -0.39 &       0.15 &   51 \\
 33107013 &    3507 &          36 &    3577 &          43 &  -0.10 &       0.11 &  100 \\
557905051 &    3618 &          38 &    3626 &          60 &  -0.15 &       0.14 &   55 \\
\hline
\end{tabular}
    \label{tab:all}
\end{table}

\bibliography{sample63}{}
\bibliographystyle{aasjournal}
\end{document}